\documentclass[a4paper,prb,showpacs,preprintnumbers,amsmath,amssymb]{revtex4}
\usepackage{bm}
\newcommand{\be}{\begin{equation}}
\newcommand{\ee}{\end{equation}}
\newcommand{\bea}{\begin{eqnarray}}
\newcommand{\eea}{\end{eqnarray}}

\begin {document}
\title{Effective Hamiltonian for a Half-filled Hubbard Chain
with Alternating On-site Interactions}
\author{Paata Kakashvili \dag \, and George I. Japaridze \ddag}
\affiliation{\dag \, Institute of Theoretical Physics, Chalmers University
of Technology and G\"oteborg University,\\
SE 412 96 G\"oteborg,
Sweden}
\affiliation{\ddag \, Andronikashvili Institute of Physics, Georgian Academy of Sciences,\\ 
Tamarashvili str. 6, 01  77, Tbilisi, Georgia}

\begin{abstract}

We derive an effective spin Hamiltonian for the one-dimensional half-filled 
Alternating Hubbard model in the limit of strong on-site repulsion. We
show that the effective Hamiltonian is a spin $S=1/2$ Heisenberg
chain with asymmetric next-nearest-neighbor (nnn) exchange.  

\end{abstract}

\maketitle

\section{Introduction}

During the last two decades the correlation induced metal-insulator (Mott)
transition has been  one of the challenging problems in condensed
matter physics \cite{MI-Tran_Books_90}. In most cases the breaking of
spatial symmetry is a prerequisite for a Mott insulator
\cite{Lee}. The undoped high-Tc copper-oxide materials are famous
example of such Mott insulators \cite{Dagotto}. However, the
one-dimensional Hubbard model \cite{Hubbard63}
\be\label{HamiltonianHubMod}
{\cal H}_{Hub} = t\sum_{i,j,\alpha}N_{i,j}c^{\dag}_{i,\alpha}c_{j,\alpha}
+ U \sum_{i}n_{i,\uparrow} n_{i,\downarrow}
\ee
at half-filling represents a case where dynamical generation of a
charge gap is not connected with the breaking of a discrete symmetry
\cite{LiebWu}. In Eq. (\ref{HamiltonianHubMod})  we have used
standard notations: $n_{i,\alpha}=c^{\dag}_{i,\alpha}c_{i,\alpha}$,
$U$ is the Hubbard on-site repulsive interaction, and  $N_{i,j}=1$ if 
$i$ and $j$ are labels for neighboring sites and equals zero
otherwise. Thus the kinetic part represents hops between neighboring
sites and the interaction part gives contributions only from electrons
on the same site. At half-filling the exact solution shows a uniform
ground state with exponentially suppressed density correlations
\cite{EsslerFrahm99} and gapless $SU(2)$ symmetric spin degrees of
freedom \cite{FrahmKorepin90}. This is in agreement with the large $U$
expansion result, that at $U \gg |t|$ the model
(\ref{HamiltonianHubMod}) is equivalent to the effective spin $S=1/2$
Heisenberg antiferromagnet Hamiltonian \cite{Anderson59},
\begin{equation}
{\cal H}_{\it Heis}= J\sum_{i} {\bf S}_{i} \cdot {\bf S}_{i+1}
+ J^{\prime}\sum_{i} {\bf S}_{i} \cdot {\bf S}_{i+2}\, ,
\label{SpinHamHM}
\end{equation}
where $J=4t^{2}/U\left(1-4t^{2}/U^{2}\right)$,
$J^{\prime}=4t^{4}/U^{3}$ in the forth-order perturbation. Several
very elegant mathematical tools have been developed for calculation of
the effective spin-chain Hamiltonians in higher orders
\cite{Takahashi77,MacDonald88,Datta99}. However, in agreement with the
exact solution \cite{LiebWu}, higher order terms are irrelevant and
the ground state remains featureless.

Discussions of the Mott-Hubbard transition within the framework of
Hubbard type models are generally restricted to lattices of
equivalent sites. However, the less common case where the spatial
invariance of the system is broken via the introduction of two types of
atoms, say  ``anions'' and ``cations'', has  attracted much recent
interest. In the  simplest case of a two-site ionic generalization of
the Hubbard model we obtain a Hamiltonian of the 
two-band Hubbard model \cite{Emery87}
\bea\label{AIHM_Hamiltonian}
{\cal H}_{} & = &  t\sum_{i,j,\alpha}N_{i,j}c^{\dag}_{i,\alpha}c_{j,\alpha}
 + \frac{\Delta}{2}\sum_{i,\alpha}\left(n_{2i,\alpha} - n_{2i+1,\alpha}
 \right)\nonumber\\
&+& U_{e} \sum_{i}n_{2i,\uparrow} n_{2i,\downarrow} +
U_{o} \sum_{i}n_{2i+1,\uparrow} n_{2i+1,\downarrow} \, .
\eea
In the limit $\Delta=0$ and $U_{e}=U_{o}=U$
Eq. (\ref{AIHM_Hamiltonian}) reduces to the ordinary Hubbard
model. The limit $U_{e}=U_{o}=U$, $\Delta \neq 0$ is called the Ionic
Hubbard Model (IHM) \cite{Nagaosa86}. In this paper we study the
Alternating Hubbard model (AHM) \cite{Japaridze93} obtained from
Eq. (\ref{AIHM_Hamiltonian}) in the limit $\Delta=0$.

The IHM and the AHM describe two different ways how the lattice
invariance can be broken. In the IHM the broken translational symmetry
is traced already by non-interacting electrons, via the {\em
  single-electron} potential energy difference between neighboring
sites ($\Delta$). In marked contrast with the IHM the
lattice unit in the AHM  is doubled {\em dynamically}, via the different {\em
  two-electron} on-site repulsion energy on even ($U_{e}$) and odd
($U_{o}$) sites.

Current interest in the study of the Mott transition in 1D models with broken
translational symmetry was triggered by the bosonization analysis of
the Ionic Hubbard Model (IHM) by Fabrizio, Nersesyan and Gogolin (FGN)
\cite{Fabrizio99}. At $U$ = 0 the IHM is a regular band insulator with
long-range ordered charge-density-wave (CDW), while in the
strong-coupling limit, at $U \gg t,\Delta$ it is a Mott insulator with
a charge gap and gapless spin sector.  The spin sector is also given
by the same Heisenberg chain  (\ref{SpinHamHM}), but with slightly renormalized
coupling constants \cite{Nagaosa86} 
\be\label{JCouplingsIHM}
J=\frac{4t^2}{U}\left[\frac{1}{1-\lambda^2}
-\frac{4t^2}{U^2}\frac{(1+4\lambda^2 
-\lambda^4)}{(1-\lambda^2)^3}\right]\, ,\quad 
J^{\prime}
=\frac{4t^4}{U^3}\frac{(1+4\lambda^2-\lambda^4)}{(1-\lambda^2)^3} \, , 
\ee
where $\lambda=\Delta/U$.

The bosonization analysis shows that the CDW-Mott insulator transition
has a complicated two-step nature \cite{Fabrizio99}: With increasing $U$
there is first an {\it Ising type transition} from a CDW band phase
into a LRO dimerized phase at $U^c_{ch}$. With further increase of
the Hubbard repulsion, at $U^{c}_{sp}>U^{c}_{ch}$ a continuous {\it
  Kosterlitz-Thouless transition} takes place from the dimerized into
a MI phase. The charge gap vanishes {\it
  only} at $U=U^c_{ch}$, while the spin sector is gapless for
$U>U^{c}_{sp}$. This phase diagram was later confirmed by various
numerical and analytical studies
\cite{Aligia01,KSJB03,Gros03,Manmana03,Nakamura04}. Moreover, these numerical
studies reveal a rather complex nature of the strong-coupling,
$U \gg \Delta,t$, insulating phase of the IHM. Although the spin sector
of the IHM in the strong-coupling limit is qualitatively similar to
that of the ordinary Hubbard model, 
the ground state is characterized by a LRO CDW  pattern and
therefore shows broken translational symmetry
\cite{KSJB03,Gros03,Manmana03}. 

In this paper we derive the effective spin Hamiltonian in the strong
on-site repulsion limit $U_{e},U_{o} \gg t$ of the AHM. As we show
below, in marked contrast with the ordinary Hubbard and the Ionic
Hubbard model, in the case of AHM the effective spin Hamiltonian is not
translational invariant and is given by a frustrated Heisenberg
chain with alternating next-nearest-neighbor exchange
\be\label{SpinHamiltonianSH}
{\cal H}_{\it eff}= J\sum_{i} {\bf S}_{i} \cdot {\bf S}_{i+1} +
\sum_{i}(J^{\prime} - (-1)^{i}\delta J^{\prime}) {\bf S}_{i} \cdot
{\bf S}_{i+2} \, , 
\ee
where
\begin{eqnarray}
J & = & 2t^{2}(\frac{1}{U_{o}}+\frac{1}{U_{e}}) -
2t^{4}(\frac{3}{U^{3}_{o}}+\frac{3}{U^{3}_{e}} 
+\frac{1}{U^{2}_{o}U_{e}}+\frac{1}{U_{o}U^{2}_{e}}), \\
J^{\prime}  & = & 2t^{4} (\frac{1}{U^{3}_{o}} +\frac{1}{U^{3}_{e}})\, , \qquad
\delta J^{\prime} = 2t^{4} \frac{U_{o}-U_{e}}{U^{2}_{o}U^{2}_{e}}\, .
\end{eqnarray}
The obtained effective Hamiltonian is the $S=\frac{1}{2}$ shark-tooth
Hamiltonian  \cite{NakamuraKubo}, the limiting case of which
($U_{e}=U_{o}=U$) is a well-known result for the Hubbard model.

\section{The strong-coupling expansion approach}

In this paper we apply the method developed by MacDonald, Girvin
and Yoshioka in the case of the ordinary Hubbard chain \cite{MacDonald88}
to obtain the effective spin Hamiltonian for the one-dimensional {\em
  Alternating Hubbard model} given by the Hamiltonian ${\cal H}_{AHM}
=\hat{T}+ \hat{V}$, where
\begin{eqnarray}
\hat{T} & = & t\sum_{i,j,\alpha}N_{i,j}
c^{\dag}_{i,\alpha}c_{j,\alpha}\\
\hat{V} & = & U_{e} \sum_{i}n_{2i,\uparrow} n_{2i,\downarrow} +
U_{o} \sum_{i}n_{2i+1,\uparrow} n_{2i+1,\downarrow} \, .
\label{HamiltonianAltHubMod}
\end{eqnarray}

In what follows we consider the strong coupling limit, assuming
$U_{e}>U_{o}\gg |t|$. Contrary to the case of the ordinary Hubbard
model where subbands can be classified by the total number of
double-occupied states (doublons) $N_{d}$, in the case of AHM we have
to deal with a system, where each band is characterized by two
different numbers: the number of doubly occupied sites in even and
odd sublattices, denoted by $N_{de}$ and $N_{do}$ respectively. The
hopping term mixes states from these subbands. The "unmixing" of the
AHM subbands can be achieved by introducing suitable linear combinations of the
uncorrelated basic states.  The ${\cal S}$ matrix for this
transformation, and the transformed Hamiltonian, $$
{\cal H}_{eff} = e^{{\it i}{\cal S}} {\cal H}_{AHM} e^{{-\it i}{\cal S}}\, ,
$$ are generated by an iterative procedure, which results in an expansion
in powers of the hopping integral $t$ divided by on-site energies
$U_e$ and/or $U_o$. 

This expansion is based on a separation of the kinetic part of the
Hamiltonian into three terms: $T_{1}$ which increases the number of
doubly occupied sites by one, $T_{-1}$ which decreases the number 
of doubly occupied sites by one and $T_{0}$ which leaves the
number of doubly occupied sites unchanged. In addition, in the
case of the Alternating Hubbard model each of these terms splits into
several different terms, depending on whether corresponding hopping
process takes place from even to odd site or vice versa.

In particular, we split the  $T_{0}$ term into three separate hopping
processes: 
\begin{equation}
T_{0}=T^{0}_{0}+T^{de}_{0}+T^{do}_{0}\, ,
\end{equation}
where
\begin{equation} \label{T01}
T^{0}_{0} = t \sum_{i,j,\alpha} N_{i,j}h_{i,-\alpha}
c^{\dag}_{i,\alpha}c_{j,\alpha}h_{j,-\alpha},
\end{equation}
is a "hole" hopping term,
\begin{equation} \label{T03}
T^{de}_{0}=t\sum_{i,j,\alpha}N_{2i,j}n_{2i,-\alpha}
c^{\dag}_{2i,\alpha}c_{j,\alpha}n_{j,-\alpha},
\end{equation}
is a "pair" hopping term, which hops a pair to even site and
\begin{equation} \label{T02}
T^{do}_{0}=t\sum_{i,j,\alpha}N_{2i+1,j}n_{2i+1,-\alpha}
c^{\dag}_{2i+1,\alpha}c_{j,\alpha}n_{j,-\alpha},
\end{equation}
is a "pair" hopping term, which hops a pair to odd site. Here
$h_{i,\alpha}=1-n_{i,\alpha}$.

The term which increases the number of doubly
occupied sites by one is also separated into two terms
$T_{1}=T^{e}_{1}+T^{o}_{1}$, where
\begin{equation} \label{Tb1}
T^{e}_{1}=t\sum_{i,j,\alpha}N_{2i,j}n_{2i,-\alpha}
c^{\dag}_{2i,\alpha}c_{j,\alpha}h_{j,-\alpha},
\end{equation}
is the term which increase the number of doubly occupied sites by one
on the sublattice of even sites and
\begin{equation} \label{Ta1}
T^{o}_{1}=t\sum_{i,j,\alpha}N_{2i+1,j}n_{2i+1,-\alpha}
c^{\dag}_{2i+1,\alpha}c_{j,\alpha}h_{j,-\alpha},
\end{equation}
is the term which increase the number of the doubly occupied sites by
one on the sublattice of odd sites.

Similarly, the term which decreases the number of doubly occupied
sites by one is also separated into two terms
$T_{-1}=T^{e}_{-1}+T^{b}_{-1}$, where 
\begin{equation} \label{Tb-1}
T^{e}_{-1}=t\sum_{i,j,\alpha}N_{i,2j}h_{i,-\alpha}
c^{\dag}_{i,\alpha}c_{2j,\alpha}n_{2j,-\alpha}
\end{equation}
and
\begin{equation} \label{Ta-1}
T^{o}_{-1}=t\sum_{i,j,\alpha}N_{i,2j+1}h_{i,-\alpha}
c^{\dag}_{i,\alpha}c_{2j+1,\alpha}n_{2j+1,-\alpha},
\end{equation}
are the terms which decrease the number of doubly occupied sites by
one on the sublattice of even and odd sites, respectively.

One can easily check the following commutation relations:
\begin{equation} \label{com}
[\hat{V},T^{q}_{m}]= (m+\delta_{m,0}) \Lambda_{q} T^{q}_{m},
\end{equation}
where
\begin{equation}
\Lambda_{q}= \left\{
\begin{array}{c}
U_{q},\hspace{1.7cm} q = e,o \\
(U_{e}-U_{o}),\hspace{0.4cm} q = de \\
(U_{o}-U_{e}),\hspace{0.4cm} q = do\\
0,\hspace{1.7cm} q = 0 \end{array} \right. \, .
\end{equation}
We must emphasize that the Eq.~(\ref{com}) is true in all cases
where allowed hopping processes connect only sublattices with
different on-site repulsion.

Let us now start to search for such a unitary transformation ${\cal
  S}$, which eliminates hops between states with different numbers of
  doubly occupied sites: 
\begin{equation} \label{trans}
{\cal H}'=e^{i{\cal S}} {\cal H} e^{-i{\cal S}} ={\cal H} + [i{\cal
  S},{\cal H}]+ \frac{1}{2}[i{\cal S},[i{\cal S},{\cal H}]]+...\, . 
\end{equation}
We follow the recursive scheme \cite{MacDonald88} which allows to determine a transformation which has the requested property to any desired order in $t/U$ where $U=(U_{e}+U_{o})/2$.  To proceed further we define:
\begin{equation}
T^{(k)}\left[\{a\},\{m\}\right]=T^{a_{1}}_{m_{1}}T^{a_{2}}_{m{2}}
\ldots T^{a_{k}}_{m_{k}}.
\end{equation}
Using Eq.~(\ref{com}) we can write
\begin{equation}
\left[\hat{V},T^{(k)}[\{a\},\{m\}]\right]=\sum_{i=1}^{k}\Lambda_{a_{i}}
(m_{i}+\delta_{m_{i},0})T^{(k)}[\{a\},\{m\}].
\end{equation}
${\cal H}^{\prime (k)}$ contains terms of order $t^{k}$, denoted by
${\cal H}^{\prime [k]}$, which couple states in different
subspaces. By definition $[V,{\cal H}'^{[k]}]\not=0$ and ${\cal
  H}^{\prime [k]}$ can be expressed in the following way
\begin{equation}
{\cal H}^{\prime [k]}=\sum_{\{a\}}\sum_{\{m\}} C^{(k)}_{\{a\}}(\{m\})
T^{(k)}[\{a\},\{m\}], \hspace{1cm} \sum_{i=1}^{k} m_{i}\not=0.
\end{equation}
If in each 'k'-th order step, we choose ${\cal S}^{(k)}={\cal
  S}^{(k-1)} + {\cal S}^{[ k]}$, 
where ${\cal S}^{[ k]}$ is the solution of the following equation 
\begin{equation}
[i{\cal S}^{[ k]},V]=- {\cal H}^{\prime [k]}
\end{equation}
and therefore equals
\begin{equation}
{\cal S}^{[ k]}=-i\sum_{a,\{m\}} \frac{C^{(k)}_{\{a\}}(\{m\})}
{\sum_{i=1}^{k} \Lambda_{a_{i}}
(m_{i}+\delta_{m_{i},0})}T^{(k)}[\{a\},\{m\}], \hspace{1cm}
\sum_{i=1}^{k} m_{i}\not= 0 \, ,
\end{equation}
then the transformed Hamiltonian
\begin{equation}
{\cal H}^{\prime (k+1)}=e^{i{\cal S}^{(k)}}{\cal H}e^{-i{\cal S}^{(k)}}
\end{equation}
contains terms of order $t^{k}/U^{k-1}$ which commute with
the unperturbed Hamiltonian and mix states within each subspace only.

\section{The Hubbard operators}

To treat correlations properly, it is important to know whether at
the beginning or at the end of hopping process a particular site is
doubly occupied or not. The introduction of so-called Hubbard
operators \cite{Hubbard67} provides us with the tool necessary for
such a full description of the local environment.  The
$X_{j}^{ab}$-operator is determined on each site of the lattice and
describes all possible transitions between the local basis states:
unoccupied $\mid 0 \rangle$, single occupied with "up"-spin $\mid +
\rangle$  and "down"-spin $\mid -\rangle$ 
and double occupied $\mid 2 \rangle$. The original electron creation
(annihilation) operators can be expressed in terms of the Hubbard
operators in the following way: 
\begin{equation}
c^{\dagger}_{i,\alpha}=X^{\alpha 0}_{i} + \alpha X^{2 -\alpha}_{i}
\hspace{1cm} c_{i,\alpha}=X^{0\alpha}_{i} + \alpha X^{-\alpha2}_{i}\, .
\end{equation}
Correspondingly, in terms of creation (annihilation) operators the
Hubbard operators have the form:
\begin{eqnarray} \label{HubOp}
X_{i}^{\alpha 0} &=& c^{\dagger}_{i,\alpha}(1-n_{i,-\alpha}),
\hspace{1.8cm} X_{i}^{2\alpha}=-\alpha 
c^{\dagger}_{i,-\alpha}n_{i,\alpha}, \nonumber \\
X_{i}^{\alpha -\alpha}& = & c^{\dagger}_{i,\alpha} c_{i,-\alpha},
\hspace{2.9cm} X_{i}^{20} =-\alpha
c^{\dagger}_{i,-\alpha}c^{\dagger}_{i,\alpha}, \\
X_{i}^{00} &=& (1-n_{i,\uparrow}) (1-n_{i,\downarrow}),
\hspace{1cm} X_{i}^{22} =
n_{i,\uparrow}n_{i,\downarrow} \nonumber \\
X_{i}^{\alpha\alpha}& = & n_{i,\alpha} (1-n_{i,-\alpha}).
\nonumber
\end{eqnarray}
The Hubbard operators which contain even (odd) number of electron
creation and annihilation operators are Bose-like (Fermi-like)
operators. They obey the following on-site multiplication rules
$X^{pq}_{i}X^{rs}_{i}=\delta_{q,r}X^{ps}_{i}$ and commutation
relations:
\begin{equation}
[X^{pq}_{i},X^{rs}_{j}]_{\pm}=\delta_{ij}(\delta_{qr}X^{ps}_{j}
\pm \delta_{ps}X^{rq}_{j}),
\end{equation}
where the upper sign corresponds to the case when both operators are
Fermi-like, otherwise the lower sign should be adopted.

It is straightforward to obtain that
\begin{eqnarray}\label{T00withX}
T^{0}_{0}&=&t\sum_{i,j}\sum_{\alpha}N_{i,j} X_{i}^{\alpha 0}X_{j}^{0\alpha}, \\
T^{do}_{0}&=&t\sum_{i,j}\sum_{\alpha}N_{2i+1,j} X_{2i+1}^{2-\alpha }X_{j}^{-\alpha 2}, \qquad
T^{de}_{0}=t\sum_{i,j}\sum_{\alpha}N_{2i,j} X_{2i}^{2-\alpha} X_{j}^{-\alpha 2}  \\
T^{o}_{1}&=&t\sum_{i,j}\sum_{\alpha}\alpha N_{2i+1,j} X_{2i+1}^{2-\alpha}X_{j}^{0\alpha}, \qquad
T^{e}_{1}=t\sum_{i,j}\sum_{\alpha}\alpha N_{2i,j} X_{2i}^{2-\alpha}X_{j}^{0\alpha} \\
T^{o}_{-1}&=&t\sum_{i,j}\sum_{\alpha}\alpha N_{i,2j+1}X_{i}^{\alpha 0}X_{2j+1}^{-\alpha 2}
\qquad
T^{e}_{-1}=t\sum_{i,j}\sum_{\alpha} \alpha N_{i,2j} X_{i}^{\alpha 0}X_{2j}^{-\alpha 2}
\label{TowithX}
\end{eqnarray}

One can easily find that the spin $S=1/2$ operators can be
rewritten in terms of the $X$-operators in the following way
\be\label{SpinXoperators}
S^{+}_{i} = c^{\dagger}_{i,\uparrow} c_{i,\downarrow} =
X^{+-}_{i}\, ,\hspace{0.3cm} S^{-}_{i} = c^{\dagger}_{i,\downarrow}
c_{i,\uparrow} = X^{-+}_{i}\, ,\hspace{0.3cm}
S^{z}_{i}= \frac{1}{2}(X^{++}_{i} - X^{--}_{i})\, .
\ee
\section{Effective Hamiltonian in the half-filled band case}

In what follows we focus on the case of the half-filled band. In this particular case the minimum
of the interacting energy is reached in the subspace with one electron per each site. Therefore, no hops are possible without increasing the number of doubly occupied sites and for any
state in this subspace $\mid \Psi_{LS} \rangle$
\begin{equation}\label{Elimination}
T^{e}_{-1} \mid \Psi_{LS} \rangle=0 \hspace{1cm} T^{o}_{-1} \mid
\Psi_{LS} \rangle=0 \hspace{1cm} T_{0} \mid \Psi_{LS} \rangle=0\, .
\end{equation}
Equation (\ref{Elimination}) may be generalized to higher orders
\begin{equation}\label{EliminationHigherOrders}
T^{k}[m] \mid \Psi_{LS} \rangle=0 \, ,
\end{equation}
if
\begin{equation}
M_{n}^{k}[m] \equiv  \sum_{i=n}^{k} m_{i} < 0\,
\end{equation}
for at least one value of n. Equation (\ref{EliminationHigherOrders}) can be used to eliminate many terms
from the expansion for ${\cal H}^{\prime}$ in the minimum $\langle \hat{V} \rangle$
subspace. Thus, in the fourth order of $\hat{T}$, the perturbed
Hamiltonian has the form:
\begin{eqnarray}\label{EffectHamInToper}
{\cal H}'^{(4)} &=& -\frac{1}{U_{o}}T^{o}_{-1}T^{o}_{1}
-\frac{1}{U_{e}}T^{e}_{-1}T^{e}_{1}
\nonumber \\
&-&\frac{1}{U^{3}_{o}}T^{o}_{-1}T^{0}_{0}T^{0}_{0}T^{o}_{1}
-\frac{1}{U^{2}_{o}U_{e}}T^{e}_{-1}T^{de}_{0}T^{0}_{0}T^{o}_{1}
-\frac{1}{U_{o}U^{2}_{e}}T^{e}_{-1}T^{0}_{0}T^{de}_{0}T^{o}_{1}
-\frac{1}{U^{2}_{o}U_{e}}T^{o}_{-1}T^{do}_{0}T^{de}_{0}T^{o}_{1}
\nonumber \\
&-&\frac{1}{U^{3}_{e}}T^{e}_{-1}T^{0}_{0}T^{0}_{0}T^{e}_{1}
-\frac{1}{U_{o}U^{2}_{e}}T^{o}_{-1}T^{do}_{0}T^{0}_{0}T^{e}_{1}
-\frac{1}{U^{2}_{o}U_{e}}T^{o}_{-1}T^{0}_{0}T^{do}_{0}T^{e}_{1}
-\frac{1}{U_{o}U^{2}_{e}}T^{e}_{-1}T^{de}_{0}T^{do}_{0}T^{e}_{1}
\nonumber \\
&-&\frac{1}{2U^{3}_{o}}T^{o}_{-1}T^{o}_{-1}T^{o}_{1}T^{o}_{1}
+\frac{1}{U^{3}_{o}}T^{o}_{-1}T^{o}_{1}T^{o}_{-1}T^{o}_{1}
+\Big(\frac{1}{2U^{2}_{o}U_{e}}+\frac{1}{2U^{2}_{e}U_{o}}\Big)T^{o}_{-1}T^{o}_{1}T^{e}_{-1}T^{e}_{1}
\nonumber\\
&+&\frac{1}{U^{3}_{e}}T^{e}_{-1}T^{e}_{1}T^{e}_{-1}T^{e}_{1}
-\frac{1}{2U^{3}_{e}}T^{e}_{-1}T^{e}_{-1}T^{e}_{1}T^{e}_{1}
+\Big(\frac{1}{2U^{2}_{o}U_{e}}+\frac{1}{2U^{2}_{e}U_{o}}\Big)T^{e}_{-1}T^{e}_{1}T^{o}_{-1}T^{o}_{1}
\nonumber\\
&-&\frac{1}{U_{o}U_{e}(U_{o}+U_{e})}
T^{o}_{-1}T^{e}_{-1}T^{o}_{1}T^{e}_{1}
-\frac{1}{U_{e}^{2}(U_{o}+U_{e})}
T^{e}_{-1}T^{o}_{-1}T^{o}_{1}T^{e}_{1}
\nonumber \\
&-&\frac{1}{U_{o}U_{e}(U_{o}+U_{e})}
T^{e}_{-1}T^{o}_{-1}T^{e}_{1}T^{o}_{1}
-\frac{1}{U_{o}^{2}(U_{o}+U_{e})}
T^{o}_{-1}T^{e}_{-1}T^{e}_{1}T^{o}_{1}
\end{eqnarray}

Using Eqs. (\ref{T00withX})-(\ref{SpinXoperators}), one can easily
rewrite the products of
$T$-terms in (\ref{EffectHamInToper}) via the Hubbard $X$
operators. Using the spin $S=1/2$ operators in
Eq. (\ref{SpinXoperators}) one obtains: 
\begin{equation}
T^{o}_{-1}T^{o}_{1} =  T^{e}_{-1}T^{e}_{1} =
-2t^{2}\sum_{i}({\bf S}_{i} \cdot {\bf S}_{i+1}-\frac{1}{4})
\end{equation}
\begin{eqnarray}
T^{o}_{-1}T^{0}_{0}T^{0}_{0}T^{o}_{1} =
T^{o}_{-1}T^{do}_{0}T^{de}_{0}T^{o}_{1} =
T^{e}_{-1}T^{0}_{0}T^{0}_{0}T^{e}_{1}=
T^{e}_{-1}T^{de}_{0}T^{do}_{0}T^{e}_{1}
=-2t^{4}\sum_{i}({\bf S}_{i} \cdot {\bf S}_{i+1}-\frac{1}{4})
\end{eqnarray}
\begin{eqnarray}
T^{e}_{-1}T^{de}_{0}T^{0}_{0}T^{o}_{1} = 
T^{o}_{-1}T^{0}_{0}T^{do}_{0}T^{e}_{1}
=-2t^{4}\sum_{i}({\bf S}_{i} \cdot {\bf S}_{i+1}-\frac{1}{4})
+2t^{4}\sum_{i}({\bf S}_{2i+1} \cdot {\bf S}_{2i+3}-\frac{1}{4})
\end{eqnarray}
\begin{eqnarray}
T^{e}_{-1}T^{0}_{0}T^{de}_{0}T^{o}_{1}=
T^{o}_{-1}T^{do}_{0}T^{0}_{0}T^{e}_{1}
=-2t^{4}\sum_{i}({\bf S}_{i} \cdot {\bf S}_{i+1}-\frac{1}{4})
+2t^{4}\sum_{i}({\bf S}_{2i} \cdot {\bf S}_{2i+2}-\frac{1}{4})
\end{eqnarray}
\begin{eqnarray}
T^{o}_{-1}T^{o}_{-1}T^{o}_{1}T^{o}_{1} =
T^{e}_{-1}T^{e}_{-1}T^{e}_{1}T^{e}_{1}
=8t^{4}\sum_{i,j\not=i-1,i,i+1}({\bf S}_{i} \cdot {\bf S}_{i+1}-\frac{1}{4})
({\bf S}_{j} \cdot {\bf S}_{j+1}-\frac{1}{4})
\end{eqnarray}
\begin{eqnarray}
T^{o}_{-1}T^{o}_{1}T^{o}_{-1}T^{o}_{1}
&=&T^{e}_{-1}T^{e}_{1}T^{e}_{-1}T^{e}_{1} =
T^{o}_{-1}T^{o}_{1}T^{e}_{-1}T^{e}_{1}
=T^{e}_{-1}T^{e}_{1}T^{o}_{-1}T^{o}_{1}\nonumber\\
&=&4t^{4}\sum_{i,j}({\bf S}_{i} \cdot {\bf S}_{i+1}-
\frac{1}{4})({\bf S}_{j} \cdot {\bf S}_{j+1}-\frac{1}{4})
\end{eqnarray}
\begin{eqnarray}
T^{o}_{-1}T^{e}_{-1}T^{o}_{1}T^{e}_{1} &=&
T^{e}_{-1}T^{o}_{-1}T^{o}_{1}T^{e}_{1}
=T^{e}_{-1}T^{o}_{-1}T^{e}_{1}T^{o}_{1}=
T^{o}_{-1}T^{e}_{-1}T^{e}_{1}T^{o}_{1}\nonumber\\
&=&4t^{4}\sum_{i,j\not=i-1,i,i+1}({\bf S}_{i} \cdot {\bf S}_{i+1}
-\frac{1}{4})({\bf S}_{j} \cdot {\bf S}_{j+1}-\frac{1}{4})
\end{eqnarray}
Therefore we find that in the $4^{th}$ order approximation the
strong-coupling effective spin Hamiltonian for the Alternating Hubbard
model is given by
\begin{eqnarray}\label{Effective_Heis_Chain_Ham}
{\cal H}_{eff}  = J\sum_{i}({\bf S}_{i} \cdot {\bf S}_{i+1}
-\frac{1}{4})
 +  J_{1}\sum_{i}( {\bf S}_{2i+1} \cdot {\bf S}_{2i+3} -\frac{1}{4})
+ J_{2} \sum_{i}({\bf S}_{2i} \cdot {\bf S}_{2i+2}-\frac{1}{4}),
\end{eqnarray}
where
\begin{eqnarray}
J & = & 2t^{2} (\frac{1}{U_{o}}+\frac{1}{U_{e}}) - 2t^{4} (\frac{3}{U^{3}_{o}} +\frac{3}{U^{3}_{e}}
+\frac{1}{U^{2}_{o}U_{e}} + \frac{1}{U_{o}U^{2}_{e}}), \\
J_{1} & = & 2t^{4}(\frac{1}{U^{3}_{o}} + \frac{1}{U^{3}_{e}} + \frac{U_{o}-U_{e}} {U^{2}_{o}U^{2}_{e}}), \\
J_{2} &= &2t^{4}(\frac{1}{U^{3}_{o}} + \frac{1}{U^{3}_{e}} - \frac{U_{o}-U_{e}} {U^{2}_{o}U^{2}_{e}}).
\end{eqnarray}
The effective Hamiltonian thus obtained is that of a frustrated Heisenberg
chain with alternating next-nearest-neighbor exchange
\cite{ChenButtneroit01}. Note that the nnn exchange is larger for two
spins separated by a site with low on-site repulsion than for spins
separated by a site with high on-site repulsion.

\section{Conclusion}

In this paper we have derived the effective spin Hamiltonian which
describes the low-energy sector of the one-dimensional half-filled
alternating Hubbard model in the limit of strong on-site
repulsion. The effective spin model is given by the Hamiltonian of the
Heisenberg chain with alternating next-nearest-neighbor exchange. This
model was intensively studied in the last few years
\cite{ChenButtneroit01,SorellaParola02,ChenButtneroit02,SenSarkar,ChenButtneroit03}.
Unfortunately, conflicting results have been reported in these
studies regarding the relevance of the alternating nnn exchange. In
some studies
\cite{ChenButtneroit01,ChenButtneroit02,ChenButtneroit03} it was
claimed that in the limit
of small frustration $J^{\prime}/J < 0.5$, the asymmetry of the nnn
exchange destabilizes the isotropic Heisenberg fixed point and leads
to a new phase with gapless excitation spectrum and vanishing
spin-wave velocity. However, other studies
\cite{SorellaParola02,SenSarkar} claim that the alternating nnn
exchange is an {\it irrelevant} perturbation. We believe that
detailed numerical studies of the phase diagram of the Alternating
Hubbard model may shed more light on this topic. The weak-coupling
bosonization analysis of the same model is in progress and will be
published separately.

\section{Acknowledgments}
It is our pleasure to thank  Henrik Johannesson, Arno Kampf,
Michael Sekania and Irakli Titvinidze for many interesting
discussions. GIJ also thanks Dionys Baeriswyl for kind hospitality
and many interesting discussions during his stay at the University
of Fribourg, where part of this work has been done. We also acknowledge
support by the SCOPES grant N 7GEP J62379.

\section*{Appendix}

Using the technique developed above it is straightforward to derive
the strong coupling effective Hamiltonian in the case of the
Alternating Ionic Hubbard model, Eq. (\ref{AIHM_Hamiltonian}), in the
limit $\Delta\neq 0$ and $U_{e} \neq U_{o} \gg \Delta, t$. 

In this case the effective Hamiltonian is also given by the Heisenberg
Hamiltonian with alternating next-nearest-neighbor exchange,
Eq. (\ref{Effective_Heis_Chain_Ham}), but with the following exchange
coupling constants: 
\begin{eqnarray}
 J  & =  & 2t^{2} \left(\frac{1}{U_{o}-\Delta} + \frac{1}{U_{e}+\Delta}\right)\nonumber \\
& - & 2t^{4} \Big[\frac{4}{(U_e + \Delta)^3} +\frac{4}{(U_o - \Delta)^3}
- \frac{1}{U_e (U_e+\Delta)^2} - \frac{1}{(U_o-\Delta)^2 U_o} \nonumber \\
& + & \frac{4(U_e+U_o)}{(U_e + \Delta)^2(U_o -  \Delta)^2}
-\frac{1}{(U_e + \Delta)^2 U_o} - \frac{1}{U_e (U_o - \Delta)^2}\nonumber \\
&-& \frac{2}{(\Delta +U_e) (U_o - \Delta) U_o}
 - \frac{2}{U_e (\Delta +U_e) (U_o - \Delta)}\Big]\, ,  \\
\nonumber\\
J_{1}&  = & 2t^{4}\Big[\frac{1}{(U_e + \Delta)^3} + \frac{1}{(U_o - \Delta)^3}
 + \frac{U_e+U_o}{(U_e + \Delta)^2 (U_o - \Delta)^2}
-\frac{2}{U_o(U_{e} + \Delta)(U_{o} - \Delta) }\Big]\, ,\\
\nonumber\\
J_{2} &= &2t^{4}\Big[\frac{1}{(U_e + \Delta)^3}+
\frac{1}{(U_o - \Delta)^3}
+\frac{U_e+U_o}{(U_e + \Delta)^2 (U_o - \Delta)^2}
-\frac{2}{U_{e}(U_{e} + \Delta) (U_o - \Delta)} \Big]\, .
\end{eqnarray}

Note, that in the case when the ionic term is large on the site with
larger on-site repulsion, i.e. $U_{e}>U_{o}$ and $\Delta > 0$, the
asymmetry in the nnn exchange in the case of AIHM
\be
\delta J^{\prime}(\Delta \neq 0)=2t^{4}
\frac{(U_{e}-U_{o})}{U_{e}U_{o}(U_{e} + \Delta)(U_{o} - \Delta)}
\ee
is larger than it is in the case of AHM
\be
\delta J^{\prime}(\Delta=0) = 2t^{4}\frac{(U_{e}-U_{o})}{U_{e}^{2}U_{o}^{2}}
\ee
for $\Delta>0$ ($\Delta \ll U_{o}, U_{e}$).

\end{document}